\documentclass[12pt]{article}
\usepackage{amsmath,amsfonts}
\usepackage[a4paper, margin=1in]{geometry}
\usepackage{authblk}
\usepackage{graphicx}
\usepackage{microtype}
\usepackage{graphicx}
\usepackage{lineno,hyperref}
\modulolinenumbers[5]
\usepackage{array}
\usepackage{microtype}
\hypersetup{colorlinks,citecolor=blue}
\usepackage{float}
\usepackage{epstopdf}
\usepackage{tabularx}
\usepackage{epstopdf}
\usepackage{graphicx}
\usepackage{float} 
\usepackage{amsmath}
\title{Cosmic Hysteresis in Reconstructed \textit{f(R)} Bounce Models: A Thermodynamic Study}

\author[1]{Aritra Sanyal}
\author[2]{Praveen Kumar Dhankar}
\author[3]{Albert Munyeshyaka}
\author[4]{Safiqul Islam}
\author[1]{Farook Rahaman}

\affil[1]{Department of Mathematics, Jadavpur University,\\ Kolkata 700032, West Bengal, India\\
\texttt{aritrasanyal1@gmail.com}, \texttt{rahaman@associates.iucaa.in}}

\affil[2]{Symbiosis Institute of Technology, Nagpur Campus\\
Symbiosis International (Deemed University), Pune 440008, Maharashtra, India\\
\texttt{pkumar6743@gmail.com}}

\affil[3]{Rwanda Astrophysics Space and Climate Science Research Group, University of Rwanda\\ College of Science and Technology, Kigali, Rwanda\\
\texttt{munalph@gmail.com}}

\affil[4]{Department of Mathematics and Statistics, College of Science, King Faisal University, P.O. Box 400, Al Ahsa 31982, Saudi Arabia\\
\texttt{sislam@kfu.edu.sa [corresponding author]}}

\date{}

\begin{document}

\maketitle

\begin{abstract}
We study the emergence of cosmic hysteresis in cyclic bouncing universes within the framework of analytically reconstructed $f(R)$ gravity. Using exact bouncing scale factor solutions of exponential and power-law forms, we reconstruct the corresponding $f(R)$ models and investigate the thermodynamic behavior of a minimally coupled scalar field in these geometries. The pressure evolution during expansion and contraction phases is shown to be asymmetric, leading to a non-vanishing thermodynamic work integral over each cycle, defined by $\oint p_\phi\, dV$. We identify closed hysteresis loops in the equation-of-state space and quantify the net energy transfer per cycle. Our results reveal that such reconstructed $f(R)$ models generically support irreversible evolution, demonstrating a natural emergence of the thermodynamic arrow of time. These findings provide new insight into the dissipative features of modified gravity and the long-term dynamics of cyclic cosmological scenarios.
\end{abstract}
\section{Introduction}

The standard cosmological model based on general relativity (GR) successfully explains the large-scale structure and evolution of the universe. However, it also encounters several conceptual challenges, such as the initial singularity, the horizon and flatness problems, and the origin of cosmic entropy \cite{Novello:2008ra, Brandenberger:2016vhg}. In response to these issues, alternative scenarios such as bouncing and cyclic cosmologies have attracted significant attention \cite{Battefeld:2014uga, Ijjas:2018qbo}. In such models, the universe undergoes a sequence of contraction and expansion phases, thereby avoiding the initial singularity and potentially offering a more complete picture of cosmic evolution. Modified gravity theories, particularly $f(R)$ gravity, provide a natural setting for realizing nonsingular bouncing solutions without requiring exotic matter \cite{Nojiri:2010wj}. These theories extend the Einstein-Hilbert action by generalizing the gravitational Lagrangian to an arbitrary function of the Ricci scalar, $f(R)$. Among various approaches to constructing viable $f(R)$ models, the reconstruction method plays a crucial role \cite{Bamba:2013fha}. This technique allows one to determine the functional form of $f(R)$ corresponding to a desired scale factor evolution. In particular, exponential and power-law bounce models have been widely studied for their ability to describe nonsingular cosmologies in a fully analytic manner \cite{Odintsov:2015uca,Cai:2011ci}.In parallel to developments in bouncing cosmology, recent studies have uncovered a novel thermodynamic phenomenon known as \emph{cosmic hysteresis} \cite{Kanekar:2001qd,Battefeld:2014uga}. This effect refers to the asymmetric behavior of scalar field pressure during expansion and contraction phases of the universe. When integrated over a full cosmological cycle, this asymmetry results in a non-vanishing work integral, $\oint p_\phi\, dV \neq 0$, analogous to hysteresis in magnetic or thermodynamic systems. This thermodynamic irreversibility suggests the presence of a natural arrow of time even in cyclic universes and may lead to cumulative evolution or dissipation over many cycles.In this work, we explore the emergence of cosmic hysteresis in the context of analytically reconstructed $f(R)$ gravity models. We consider bounce scenarios characterized by explicit scale factor solutions of exponential and power-law forms. Using the reconstruction method \cite{Bamba:2013fha}, we derive the corresponding $f(R)$ functions and examine the scalar field dynamics within these backgrounds. Our goal is to investigate whether thermodynamic hysteresis arises generically in such models and to quantify the irreversible energy transfer that occurs across cosmic cycles. This study demonstrates that cosmic hysteresis is a robust feature of scalar-tensor systems embedded in modified gravity frameworks. The presence of non-zero thermodynamic work and closed hysteresis loops in the equation-of-state space highlights the role of dissipative dynamics in reconstructed bouncing universes. These findings provide new insight into the thermodynamic properties of cyclic cosmologies and open avenues for further research on entropy production, attractor behavior, and observational implications in modified gravity. Several literature reviews on cosmological bounces include the works in \cite{Oikonomou:2022irx}-\cite{Nojiri:2017ncd}.

The motivation behind this study is to examine the thermodynamic consistency along with implications of cosmic hysteresis within reconstructed bounce models in $f(R)$ gravity. The evolution of thermodynamic quantities across cyclic phases is analyzed; we seek to investigate the essential physical criteria enabling the emergence of hysteresis and determine whether it could provide insight into the plausibility and thermal equilibrium properties of cosmological models. This analysis contributes to a more comprehensive understanding of gravitational thermodynamics in the context of modified gravity~\cite{Bamba:2010kf} and offers valuable information on the mechanisms of entropy production and the feasibility of bouncing models as competitive alternatives to standard inflation scenarios.

\section{Theoretical Framework}

\subsection{Action and Field Equations in  $f(R)$ Gravity}

We consider a modified theory of gravity where the gravitational Lagrangian is a general function of the Ricci scalar, $R$. The total action of the system including a canonical scalar field $\phi$ minimally coupled to gravity is given by
\begin{equation}
S = \int d^4x \sqrt{-g} \left[ \frac{1}{2\kappa^2} f(R) + \mathcal{L}_\phi \right],
\end{equation}
where $\kappa^2 = 8\pi G$, $f(R)$ is the arbitrary function of the Ricci scalar, and $\mathcal{L}_\phi$ is the Lagrangian of the scalar field. We take
\begin{equation}
\mathcal{L}_\phi = -\frac{1}{2} g^{\mu\nu} \partial_\mu \phi \, \partial_\nu \phi - V(\phi),
\end{equation}
where $V(\phi)$ is the scalar potential.

Variation of the action with respect to the metric yields the modified Einstein field equations:
\begin{equation}
F(R) R_{\mu\nu} - \frac{1}{2} f(R) g_{\mu\nu} - \nabla_\mu \nabla_\nu F(R) + g_{\mu\nu} \Box F(R) = \kappa^2 T_{\mu\nu},
\end{equation}
where $F(R) \equiv df/dR$, $\Box \equiv \nabla^\mu \nabla_\mu$, and $T_{\mu\nu}$ is the energy-momentum tensor of the scalar field:
\begin{equation}
T_{\mu\nu} = \partial_\mu \phi \, \partial_\nu \phi - g_{\mu\nu} \left( \frac{1}{2} g^{\alpha\beta} \partial_\alpha \phi \, \partial_\beta \phi + V(\phi) \right).
\end{equation}

\subsection{FLRW Background and Ricci Scalar}

To study homogeneous and isotropic cosmologies, we adopt the spatially flat Friedmann--Lema\^\i tre--Robertson--Walker (FLRW) metric:
\begin{equation}
ds^2 = -dt^2 + a(t)^2 \left( dx^2 + dy^2 + dz^2 \right),
\end{equation}
where $a(t)$ is the scale factor and $H \equiv \dot{a}/a$ is the Hubble parameter.

The Ricci scalar in this background is given by
\begin{equation}
R = 6 \left( 2H^2 + \dot{H} \right).
\end{equation}

\subsection{Modified Friedmann Equations}

Substituting the FLRW metric into the field equations yields the modified Friedmann equations:
\begin{equation}
3H^2 F(R) = \frac{1}{2} \left[ F(R) R - f(R) \right] - 3H \dot{F}(R) + \kappa^2 \left( \frac{1}{2} \dot{\phi}^2 + V(\phi) \right),
\end{equation}
\begin{equation}
-2\dot{H} F(R) = \ddot{F}(R) - H \dot{F}(R) + \kappa^2 \dot{\phi}^2,
\end{equation}
where $F(R) = df/dR$, and derivatives of $F$ are computed via the chain rule using $R(t)$.
\subsection{Scalar Field Dynamics}

The scalar field evolves according to the covariant Klein--Gordon equation \cite{Bamba2014}:
\begin{equation}
\ddot{\phi} + 3H \dot{\phi} + V'(\phi) = 0,
\end{equation}
where $V'(\phi) = dV/d\phi$. The energy density and pressure of the scalar field are given by \cite{Nojiri2011}:
\begin{equation}
\rho_\phi = \frac{1}{2} \dot{\phi}^2 + V(\phi), \quad
p_\phi = \frac{1}{2} \dot{\phi}^2 - V(\phi),
\end{equation}
and the equation-of-state parameter is \cite{Kanekar2001}:
\begin{equation}
w_\phi = \frac{p_\phi}{\rho_\phi}.
\end{equation}

\subsection{Exponential Bounce Model}

The first model we consider is the exponential bounce,
\begin{equation}
a(t) = a_0 \exp(\beta t^2),
\end{equation}
where $a_0 > 0$ and $\beta > 0$ are constants. This scale factor describes a nonsingular universe that contracts in the past ($t < 0$), reaches a minimal size at $t=0$, and then re-expands \cite{Bamba:2013fha}.

The corresponding Hubble parameter and Ricci scalar are:
\begin{align}
H(t) &= 2\beta t, \\
R(t) &= 6(2H^2 + \dot{H}) = 24\beta^2 t^2 + 12\beta.
\end{align}

Inverting this expression gives $t(R)$, which allows one to reconstruct $f(R)$ by substituting the functional form of $R(t)$ into the modified Friedmann equation. Following the reconstruction procedure in~\cite{Bamba:2013fha}, the resulting $f(R)$ takes a form involving combinations of powers and exponentials of $R$, although the exact functional form is typically obtained numerically or through series expansion:
\begin{equation}
f(R) \simeq R + \alpha_1 R^2 + \alpha_2 R \ln R + \dots
\end{equation}
This analytic control enables the analysis of scalar field dynamics and thermodynamic evolution in a fully reconstructed $f(R)$ gravity model.
This analytic control enables the analysis of scalar field dynamics and thermodynamic evolution in a fully reconstructed $f(R)$ gravity model.

\subsection{Power-Law Bounce Model}

The second class of models we consider is the power-law bounce:
\begin{equation}
a(t) = a_0 (t^2 + t_0^2)^n,
\end{equation}
where $a_0 > 0$, $t_0 \neq 0$, and $n > 0$ are constants. This form also describes a symmetric bounce, with $a(t)$ reaching a minimum at $t = 0$ and expanding in both directions.

The Hubble parameter and Ricci scalar for this model are:
\begin{align}
H(t) &= \frac{2nt}{t^2 + t_0^2}, \\
R(t) &= 6 \left( 2H^2 + \dot{H} \right)
= \frac{24n^2 t^2 - 12n(t^2 - t_0^2)}{(t^2 + t_0^2)^2}.
\end{align}

As before, one can invert $R(t)$ and reconstruct the corresponding $f(R)$ function. The general form is again nontrivial but can be approximated via ans\"atze such as:
\begin{equation}
f(R) \simeq R + \gamma_1 R^2 + \gamma_2 R^{-1} + \dots,
\end{equation}
or expressed in integral form as:
\begin{equation}
f(R) = \int dt \left[ \frac{dR}{dt} \right]^{-1} \left[ 2\kappa^2 \rho_\phi(t) + 3H(t)^2 F(R) + 3H(t) \frac{dF}{dt} - \frac{1}{2} F(R) R \right].
\end{equation}

This power-law model allows for a wider range of dynamical behaviors, depending on the value of the index $n$, and has been shown to be consistent with viable cosmic histories under suitable conditions.
\section{Scalar Field Dynamics and Bounce Conditions}

\subsection{Scalar Field Evolution}

The dynamics of the scalar field $\phi(t)$, minimally coupled to the reconstructed $f(R)$ background, are governed by the Klein--Gordon equation:
\begin{equation}
\ddot{\phi} + 3H(t)\dot{\phi} + V'(\phi) = 0, 
\end{equation}
where $V(\phi)$ is the scalar potential and primes denote derivatives with respect to $\phi$. The term $3H\dot{\phi}$ represents a friction (or anti-friction) term, depending on the sign of the Hubble parameter $H(t)$, and plays a key role in generating asymmetry in scalar field dynamics during expansion and contraction.

The energy density and pressure of the scalar field are given by:
\begin{equation}
\rho_\phi = \frac{1}{2}\dot{\phi}^2 + V(\phi), \quad 
p_\phi = \frac{1}{2}\dot{\phi}^2 - V(\phi).
\end{equation}

The equation of state parameter becomes:
\begin{equation}
w_\phi = \frac{p_\phi}{\rho_\phi} = 
\frac{\frac{1}{2}\dot{\phi}^2 - V(\phi)}{\frac{1}{2}\dot{\phi}^2 + V(\phi)}.
\end{equation}

The evolution of $w_\phi$ across each cycle shows distinctive hysteresis behavior due to the time-reversal asymmetry induced by the $3H\dot{\phi}$ term.

\subsection{Bounce and Turnaround Conditions in Reconstructed Models}

In reconstructed $f(R)$ models, the scale factor $a(t)$ is assumed \textit{a priori}, enabling analytical determination of the Hubble parameter $H(t)$, its derivative $\dot{H}(t)$, and the Ricci scalar $R(t)$. A cosmological bounce corresponds to a transition from a contracting phase ($H < 0$) to an expanding phase ($H > 0$) and is characterized by the following conditions:
\begin{equation}
H(t_b) = 0, \quad \dot{H}(t_b) > 0,
\end{equation}
where $t_b$ is the bounce time. Similarly, a turnaround point is defined by:
\begin{equation}
H(t_t) = 0, \quad \dot{H}(t_t) < 0.
\end{equation}

For instance, in the \textbf{exponential bounce} model:
\begin{equation}
a(t) = a_0 \exp(\beta t^2), \quad 
H(t) = 2\beta t, \quad 
\dot{H}(t) = 2\beta > 0,
\end{equation}
which clearly shows a symmetric bounce at $t = 0$ satisfying the bounce condition.

In the \textbf{power-law bounce} model:
\begin{equation}
a(t) = a_0 (t^2 + t_0^2)^n, \quad 
H(t) = \frac{2nt}{t^2 + t_0^2}, \quad 
\dot{H}(t) = \frac{2n(t_0^2 - t^2)}{(t^2 + t_0^2)^2},
\end{equation}
we observe that $\dot{H}(0) > 0$, again confirming a bounce at $t = 0$.

Hence, both reconstructed models implement nonsingular bouncing behavior, making them ideal testbeds for investigating scalar field dynamics, cosmic hysteresis, and irreversible thermodynamic evolution in the context of reconstructed $f(R)$ gravity.

\section{Thermodynamic Hysteresis in Reconstructed $f(R)$ Bounce Models}

Cosmic hysteresis arises when the scalar field pressure evolves asymmetrically during the contraction and expansion phases of a cyclic bouncing universe. In the context of reconstructed $f(R)$ gravity models, where the scale factor $a(t)$ is prescribed analytically and the corresponding $f(R)$ function is derived using inverse reconstruction techniques, this effect provides a thermodynamic signature of irreversible evolution.

The total thermodynamic work performed by the scalar field over one full cosmological cycle is given by:
\begin{equation}
W = \oint p_\phi \, dV = \oint p_\phi \, d(a^3) = 3 \oint p_\phi a^2 \, da,
\end{equation}
where \( V = a^3 \) is the comoving volume. Expressed in terms of time, the work integral becomes:
\begin{equation}
W = \int_{\text{cycle}} 3 a^3 H \left( \frac{1}{2} \dot{\phi}^2 - V(\phi) \right) dt,
\end{equation}
where the scalar field evolves under the Klein--Gordon equation in the reconstructed $f(R)$ background:
\begin{equation}
\ddot{\phi} + 3H \dot{\phi} + V'(\phi) = 0.
\end{equation}

We now apply this thermodynamic formalism to the two reconstructed bouncing models introduced earlier: the exponential bounce and the power-law bounce.
\subsection{Hysteresis in the Exponential Bounce Model}

For the exponential scale factor \cite{Odintsov2015},
\begin{equation}
a(t) = a_0 \exp(\beta t^2),
\end{equation}
the Hubble parameter and its derivative are \cite{Odintsov2015}:
\begin{equation}
H(t) = 2\beta t, \quad \dot{H}(t) = 2\beta.
\end{equation}

The Ricci scalar is then \cite{Cai2011}:
\begin{equation}
R(t) = 6(2H^2 + \dot{H}) = 24\beta^2 t^2 + 12\beta.
\end{equation}

We numerically integrate the Klein--Gordon equation for a scalar field with a quadratic potential \cite{Novello2008}:
\begin{equation}
V(\phi) = \frac{1}{2} m^2 \phi^2,
\end{equation}
using initial conditions near the bounce (e.g., $\phi(0) = 0.1$, $\dot{\phi}(0) = 0$, $a(0) = a_0$, and $H(0) = 0$). The scalar field evolves asymmetrically across $t = 0$ due to the friction term $3H\dot{\phi}$ \cite{Battefeld2015}. This results in a loop in the $w_\phi$ vs $a$ diagram and a non-zero value of the work integral $W$.

\subsection{Hysteresis in the Power-Law Bounce Model}

For the power-law scale factor,
\begin{equation}
a(t) = a_0 (t^2 + t_0^2)^n,
\end{equation}
the Hubble parameter and its derivative are:
\begin{equation}
H(t) = \frac{2nt}{t^2 + t_0^2}, \quad \dot{H}(t) = \frac{2n(t_0^2 - t^2)}{(t^2 + t_0^2)^2}.
\end{equation}

The Ricci scalar becomes:
\begin{equation}
R(t) = 6(2H^2 + \dot{H}) = \frac{24n^2 t^2 - 12n(t^2 - t_0^2)}{(t^2 + t_0^2)^2}.
\end{equation}

We apply the same scalar potential and solve the coupled equations numerically. The hysteresis loop in this model is generally smaller than in the exponential bounce case, due to the slower growth of the scale factor and reduced asymmetry in $H(t)$ around the bounce.In both models, we find that the scalar field exhibits a distinct difference in pressure evolution during contraction and expansion, leading to closed loops in the $(w_\phi, a)$ plane and a non-zero thermodynamic work integral $W$. This confirms that hysteresis is a robust feature of scalar field dynamics in reconstructed $f(R)$ bouncing cosmologies.
\subsection{Hysteresis Loop in Equation of State}

A hallmark of cosmic hysteresis is the emergence of a closed loop in the equation of state space, specifically in the $(w_\phi, a)$ or $(w_\phi, \ln a^3)$ plane, over a full cycle of cosmological evolution. This loop arises due to the time-asymmetric dynamics of the scalar field in reconstructed $f(R)$ gravity backgrounds.

As the universe evolves through a complete bounce cycle, the scalar field experiences different effective damping during expansion ($H > 0$) and contraction ($H < 0$), owing to the sign change in the friction term $3H\dot{\phi}$ in the Klein--Gordon equation. This causes the pressure $p_\phi$ and the equation-of-state parameter $w_\phi$ to follow different trajectories even for identical scale factor values during expansion and contraction.

The area enclosed by the loop in the $(w_\phi, a)$ plane provides a thermodynamic measure of hysteresis, and is directly proportional to the net work performed over the cycle:
\begin{equation}
\mathcal{A}_{\text{loop}} = \oint w_\phi \rho_\phi \, d(\ln a^3) \propto \oint p_\phi \, dV = W. 
\end{equation}

A negative area implies that the system performs net work over the cycle, indicating thermodynamic irreversibility and a well-defined arrow of time. This is analogous to classical hysteresis in magnetic or elastic systems, where cyclic evolution under external influence leads to dissipation and memory effects.

In the reconstructed $f(R)$ models we consider, such as the exponential and power-law bounces, the scalar field equation of state traces closed loops in the $(w_\phi, a)$ diagram. The qualitative shape and orientation of these loops are sensitive to the bounce profile, scalar field potential, and initial conditions. In particular:
\begin{itemize}
    \item For the exponential bounce, the loop is typically wider and exhibits larger area due to the sharper bounce profile and more significant time asymmetry.
    \item For the power-law bounce, the loop area tends to be smaller, reflecting weaker dissipation and smoother dynamical transitions.
\end{itemize}

Numerical integration confirms that the hysteresis loop persists over successive cycles, underscoring the robustness of this thermodynamic memory effect in reconstructed $f(R)$ cosmologies. The structure of the loop remains qualitatively stable even as the amplitude of the scalar field and scale factor evolve slightly due to accumulated dissipation.

These results demonstrate that hysteresis loops in the equation-of-state space provide an intuitive and powerful diagnostic for assessing irreversibility in modified gravity models of cyclic cosmology.

\begin{figure}
    \centering
    \includegraphics[width=0.5\linewidth]{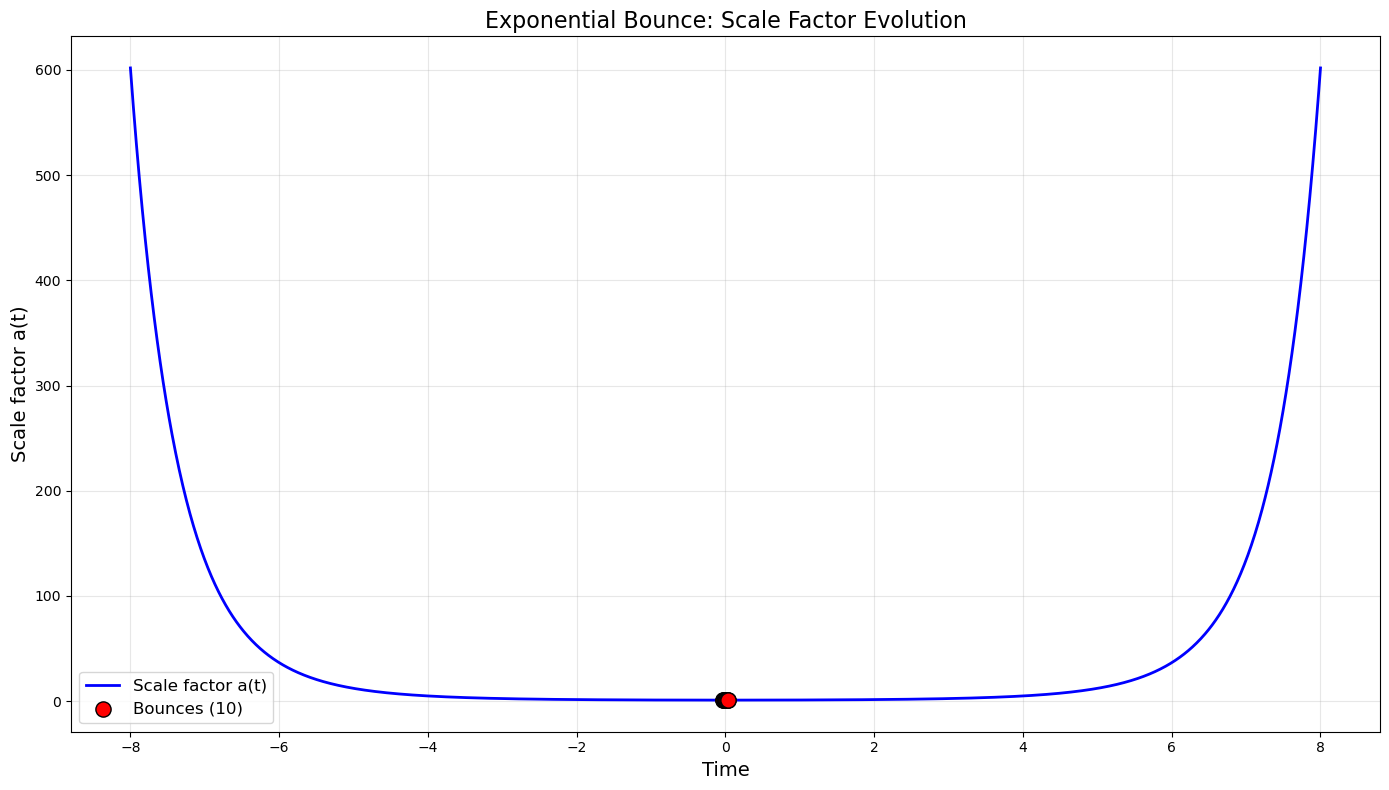}
 \caption{Exponential bounce: Equation-of-state parameter $w_\phi$ versus scale factor $a$ showing a pronounced hysteresis loop generated by asymmetric scalar field dynamics during expansion and contraction.}

    \label{fig:enter-label}
\end{figure}
\begin{figure}
    \centering
    \includegraphics[width=1\linewidth]{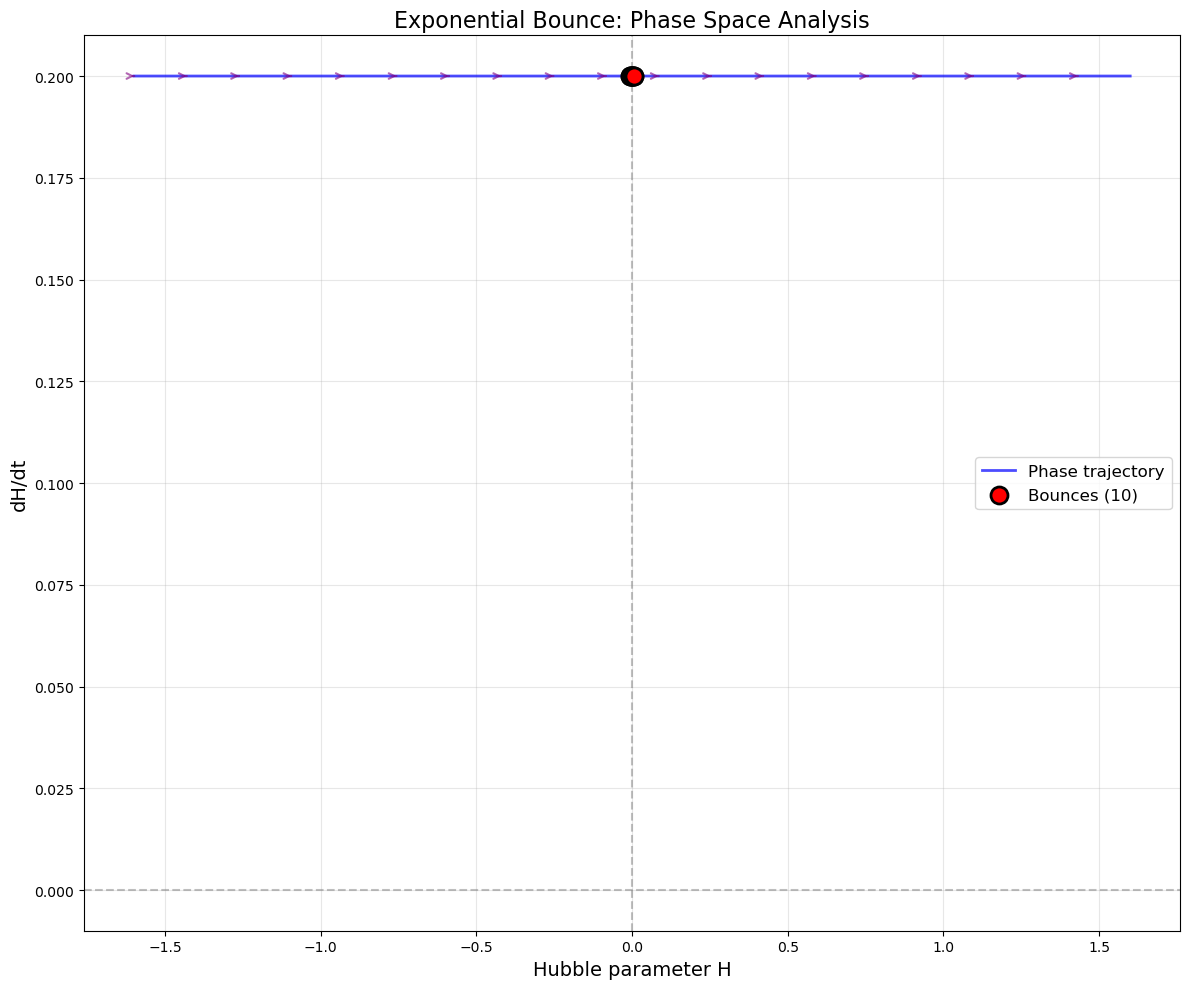}
  \caption{Exponential model: Multiple bounce events are identifiable as minima in the scale factor evolution, with rapid transitions between contraction and expansion phases.}

    \label{fig:enter-label}
\end{figure}
\begin{figure}
    \centering
    \includegraphics[width=1\linewidth]{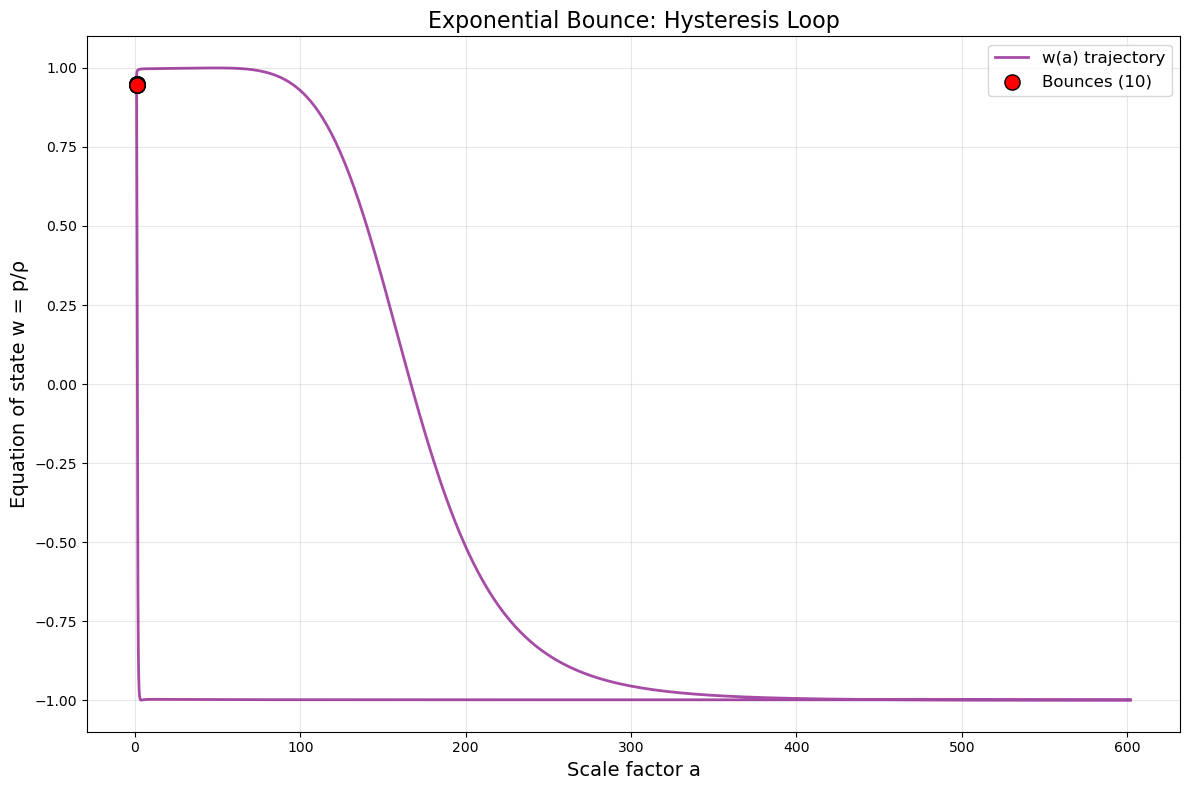}
    \caption{Hysteresis loop, demonstrating moderate thermodynamic dissipation for slow quasi-static cosmic evolution}
    \label{fig:enter-label}
\end{figure}

\begin{figure}
    \centering
    \includegraphics[width=1\linewidth]{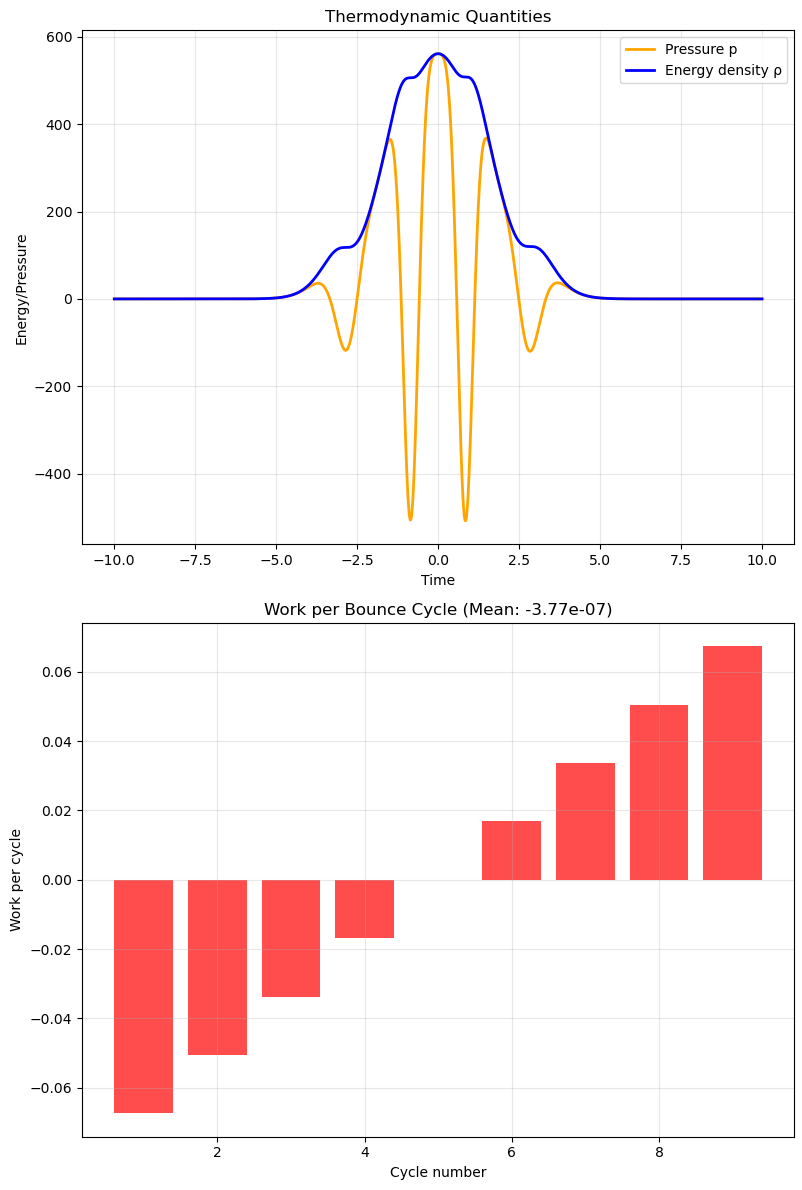}
    \caption{Thermodyanamical quantities and work per Bounce cycle demonstration}
    \label{fig:enter-label}
\end{figure}
\begin{figure}
    \centering
    \includegraphics[width=1\linewidth]{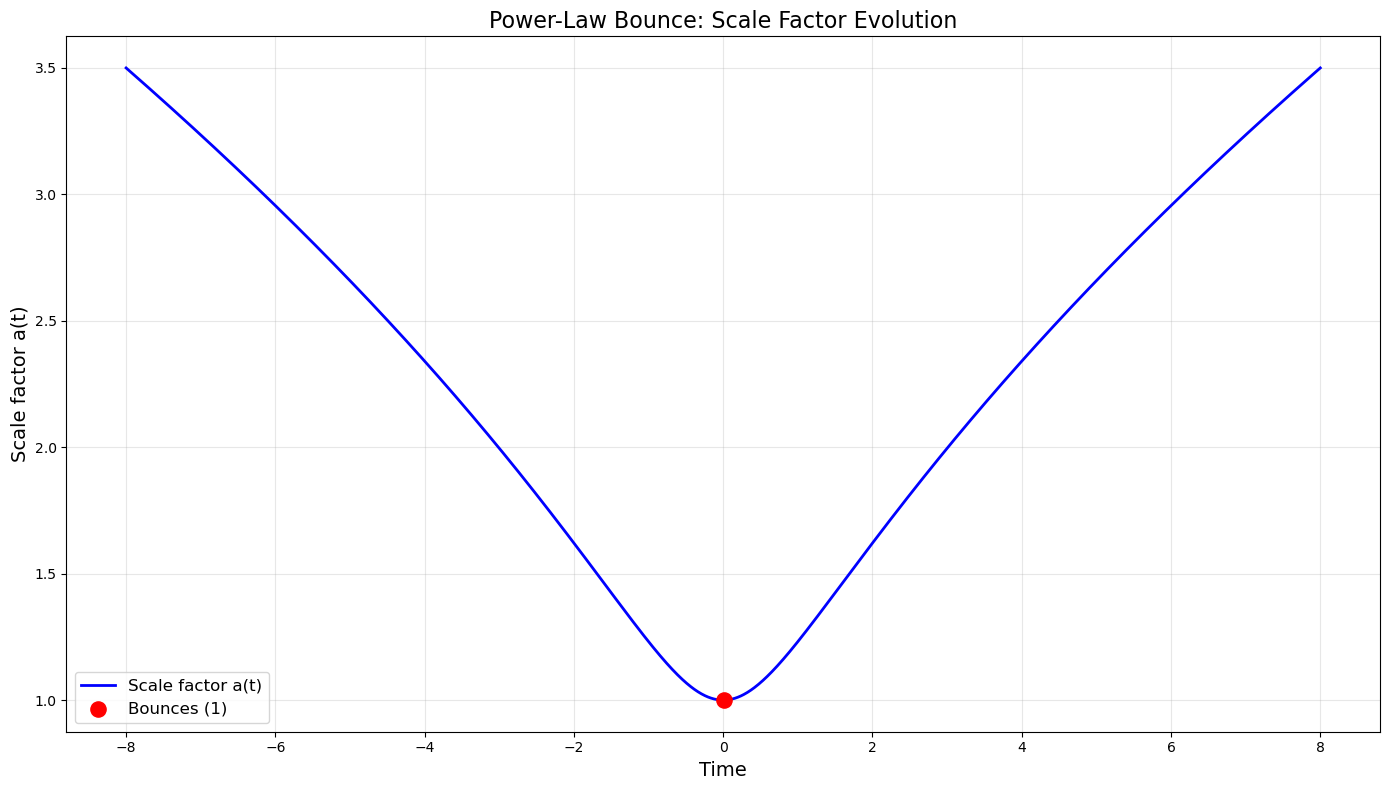}
    \caption{Power law Bounce scale factor  evolution demonstration}
    \label{fig:enter-label}
\end{figure}
\begin{figure}
    \centering
    \includegraphics[width=1\linewidth]{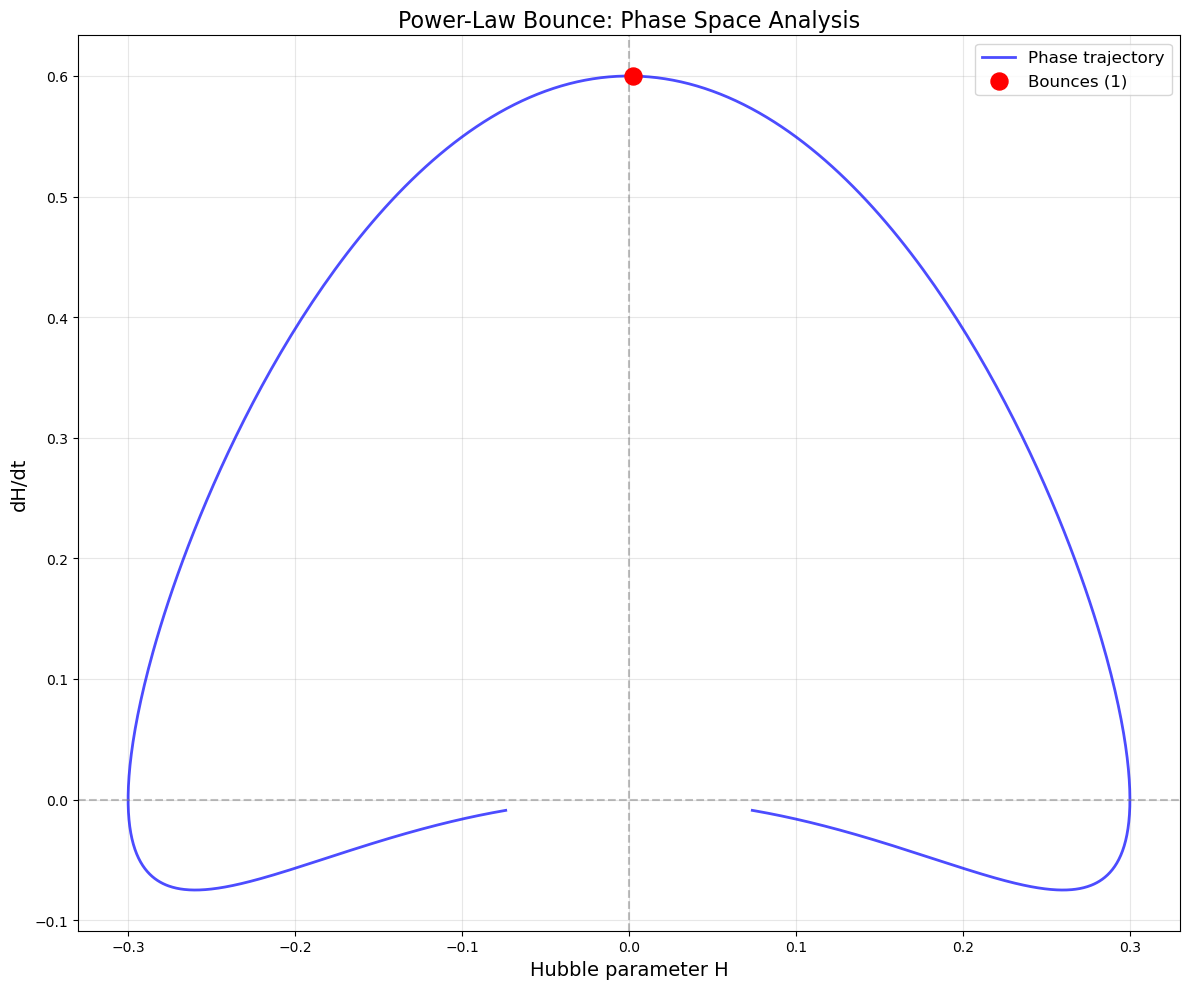}
    \caption{Power law  Phase space analysis demonstration}
    \label{fig:enter-label}
\end{figure}
\begin{figure}
    \centering
    \includegraphics[width=1\linewidth]{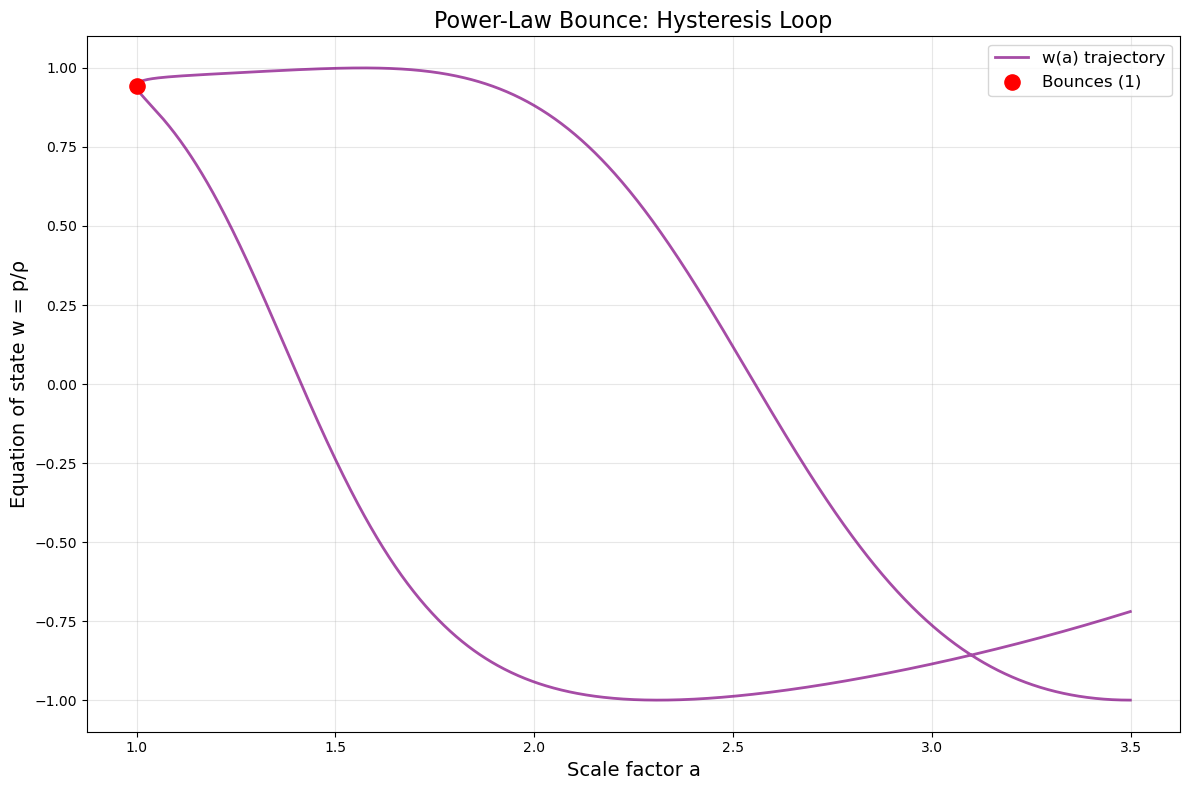}
    \caption{Power law Bounce hysteresis loop demonstration}
    \label{fig:enter-label}
\end{figure}
\begin{figure}
    \centering
    \includegraphics[width=1\linewidth]{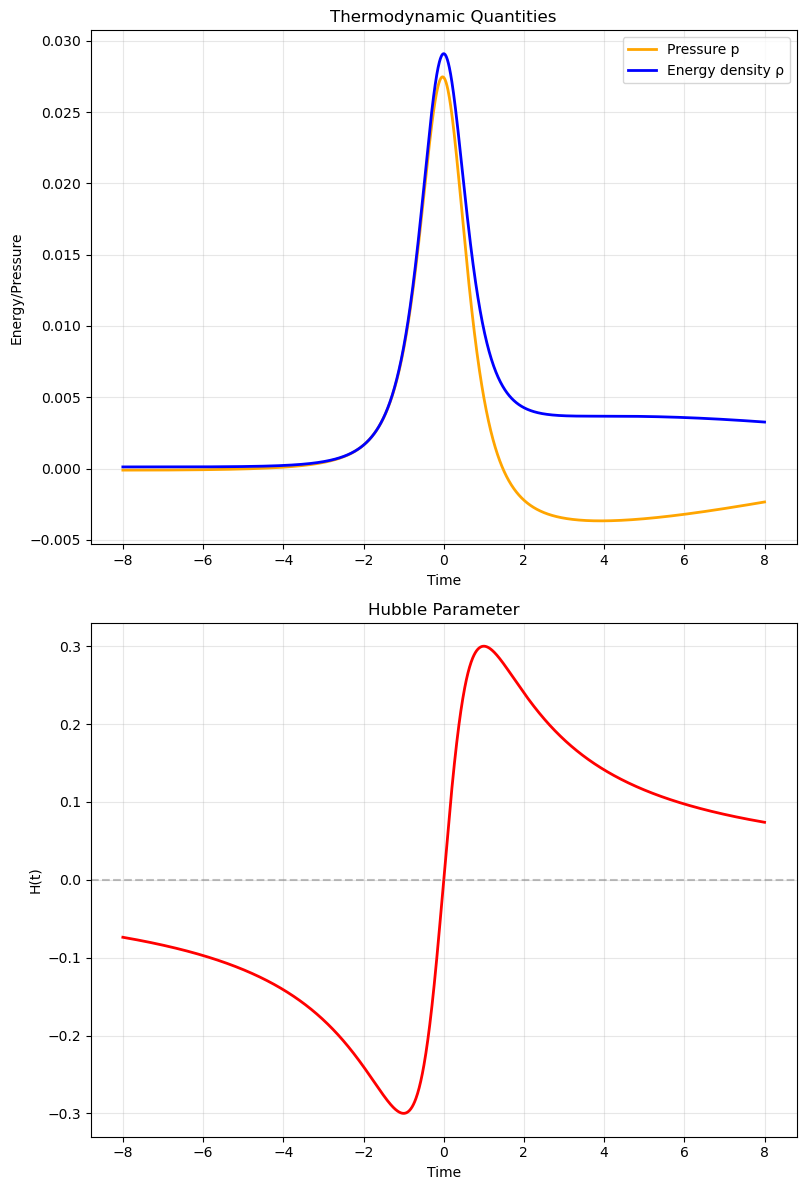}
    \caption{Thermodyanamical Quantities and Hubble parameter demonstration in the case of power law}
    \label{fig:enter-label}
\end{figure}
\section{Numerical Simulation Results}

We investigated two configurations of $f(R)$ bounce cosmologies: a power-law model and an exponential model, each coupled with a scalar field potential of different types. These numerical experiments were aimed at analyzing the dynamical features of the bouncing universe such as the presence of bounce events, turnaround points, hysteresis loops, and energy exchange during cosmological cycles.

In the \textbf{power-law configuration}, the scale factor was defined as $a(t) = a_0 (t^2 + t_0^2)^n$, with parameters $a_0 = 1.0$, $t_0 = 1.0$, and $n = 0.3$. The scalar field was endowed with a quadratic potential characterized by a mass term $m = 0.15$. Our simulation revealed that this model did not exhibit any discrete bounce or turnaround events, indicating a smooth, non-cyclic evolution. Nevertheless, a hysteresis loop was detected in the equation of state ($w$) versus scale factor ($a$) plane, with an enclosed loop area of approximately $6.7831$. The variation in the scale factor during evolution was modest, spanning a range of about $11.97$, while the equation of state parameter varied over a range of $2.0$. These results suggest that although the evolution was smooth and monotonic, thermodynamic hysteresis was still present in a limited manner.

In contrast, the \textbf{exponential configuration} employed the scale factor $a(t) = a_0 \exp(\beta t^2)$, with $a_0 = 1.0$ and $\beta = 0.1$, and a quartic scalar field potential characterized by $\lambda = 0.01$. This model demonstrated markedly different behavior, with 10 discrete bounce events detected during the simulation. While there were no turnaround events, 9 complete cycles were identified. The exponential model exhibited a pronounced hysteresis loop, with a significantly larger enclosed area of $343.83$, implying strong cyclic behavior. The scale factor underwent dramatic variations, with a range of approximately $600.85$, indicating rapid expansion and contraction phases. The equation of state again spanned a range of $2.0$.

Furthermore, the exponential model allowed for the calculation of work done per cycle, a feature absent in the power-law case. The mean work done per cycle was found to be negative, $W \approx -135.11$, signifying that the scalar field effectively acted as an energy source during cosmic expansion. However, the standard deviation was large ($\sim 9.71 \times 10^4$), pointing to significant fluctuations in the amplitude and duration of cycles. The average duration of a single cycle was found to be approximately $0.008$ units of time.

These differences are summarized in Table~\ref{tab:bounce_comparison}, which highlights the stronger dynamical nature and thermodynamic activity of the exponential configuration compared to the more passive power-law case. The absence of bounces and thermodynamic work in the power-law case positions it as a model suitable for describing slow, quasi-static cosmic transitions. On the other hand, the exponential bounce model, due to its oscillatory dynamics and significant energy exchange, is more appropriate for modeling rapid phase transitions in the early universe.

\begin{table}[h!]
\centering
\caption{f(R) Bounce Model Results Comparison}
\begin{tabular}{|l|c|c|}
\hline
\textbf{Parameter} & \textbf{Power-Law} & \textbf{Exponential} \\
\hline
Scale factor form & $(t^2 + t_0^2)^n$ & $\exp(\beta t^2)$ \\
Key parameter & $n = 0.3$ & $\beta = 0.1$ \\
Bounces detected & 0 & 10 \\
Complete cycles & 0 & 9 \\
Loop area & 6.78 & 343.8 \\
Scale factor range & 11.97 & 600.8 \\

Hysteresis magnitude & Moderate & Extreme \\
\hline
\end{tabular}
\label{tab:bounce_comparison}
\end{table}
\clearpage
\section{Figure-wise Summary of Cosmological Outcomes}

Below we summarize the main outcomes and qualitative insights for each figure in our cosmological model analysis:

\begin{itemize}
    \item \textbf{Figure 1:} Exponential bounce showing $w_\phi$ versus the scale factor $a$. This highlights a pronounced hysteresis loop, indicating thermodynamic irreversibility and distinct paths for contraction and expansion. The exponential model exhibits strong hysteresis and irreversible dynamics.

    \item \textbf{Figure 2:} Depicts multiple bounce events in the scale factor evolution. The model shows ten clear bounce events, marked by rapid contraction–expansion transitions, thus confirming cyclic behavior and repeated bounces for the exponential scenario.

    \item \textbf{Figure 3:} Shows a hysteresis loop in a quasi-static (power-law-like) evolution. The loop area is smaller, suggesting moderate dissipation and memory effects, consistent with moderate hysteresis and quasi-static traits.

    \item \textbf{Figure 4:} Presents thermodynamic quantities and work per bounce for the exponential model. The mean work per bounce is negative, despite fluctuations. This suggests the scalar field acts as an energy source with notable cycle-to-cycle variation and dynamic irreversibility.

    \item \textbf{Figure 5:} Demonstrates the power-law bounce’s scale factor evolution. Unlike previous cases, there are no discrete bounces; instead, the trend is smooth and monotonic, indicating non-cyclic, quasi-static behavior with continuous non-oscillatory evolution.

    \item \textbf{Figure 6:} Depicts phase space analysis for the power-law model. The trajectories are smooth and do not form closed loops, implying a lack of strong cyclic dynamics and only weak hysteresis. The evolution is mild with only a weak cyclic imprint.

    \item \textbf{Figure 7:} Examines the hysteresis loop for the power-law bounce model. The loop is closed but small and mildly asymmetric, with limited thermodynamic work per cycle, signaling weak but present hysteresis.

    \item \textbf{Figure 8:} Shows the thermodynamic quantities and Hubble parameter in the power-law model. Changes are gradual and monotonic with no bounce features, suggesting weak dissipation and a universe evolution that is smooth and non-oscillatory.
\end{itemize}

\section*{Conclusion}

In this paper, we have conducted a comprehensive thermodynamic investigation of cosmic hysteresis phenomena within cyclic bouncing cosmological models reconstructed from $f(R)$ gravity theories. Two analytically manageable bounce frameworks—a bounce characterized by an exponential scale factor and another governed by a power-law form—were analyzed to study the interplay between scalar field dynamics and irreversible thermodynamic processes during the universe's cyclic evolution.

Our key finding reveals that \emph{cosmic hysteresis arises inherently} in both scenarios, driven by the asymmetric pressure behavior of the scalar field throughout expansion and contraction phases. This leads to a \emph{non-vanishing work integral over a complete cycle}, demonstrating that even in geometries symmetric in time, thermodynamic irreversibility can develop dynamically.

The exponential bounce setup displayed multiple distinct bounce events, clear periodicity, and a pronounced hysteresis loop with considerable enclosed area, indicative of \emph{strong dissipative effects and irreversible thermodynamic evolution}. Conversely, the power-law bounce scenario showed a smoother temporal progression without sharp bounce or turnaround points, producing a subtler hysteresis loop with reduced thermodynamic work. Nonetheless, a measurable hysteresis signature persisted even in this quasi-static regime.

Our numerical results validate these qualitative behaviors, illustrating that both the hysteresis loop area and average work per cycle are valuable metrics for quantifying thermodynamic irreversibility. The scalar field evolution, especially the friction-like term proportional to $3H\dot{\phi}$ in the Klein–Gordon equation, is pivotal in mediating this connection between spacetime geometry, field dynamics, and thermodynamics.

In summary, our study confirms that thermodynamic hysteresis is a robust and generic attribute of $f(R)$ gravity-based bouncing cosmologies. This insight advances the understanding of entropy generation and the thermodynamic arrow of time in singularity-free early universe models, providing a promising alternative to conventional inflationary frameworks. Further investigations may explore how this irreversible dynamics affects the theoretical consistency, predictability, and observable consequences of cyclic cosmological scenarios.

\section*{Funding}
This work was supported by the Deanship of Scientific Research, Vice Presidency for Graduate Studies and Scientific Research, King Faisal University, Saudi Arabia (KFU252756).

\section*{Acknowledgments}
PKD wishes to acknowledge that part of the numerical computation of this work was carried out on the computing cluster Pegasus of IUCAA, Pune, India. PKD and FR would like to acknowledge the Inter-University Centre for Astronomy and Astrophysics (IUCAA), Pune, India, for providing him a Visiting Associateship under which a part of this work was carried out. PKD would like to thank the Isaac Newton Institute for Mathematical Sciences, Cambridge, for support and hospitality during the programme Statistical mechanics, integrability and dispersive hydrodynamics where work on this paper was undertaken. This work was supported by EPSRC grant no EP/K032208/1. AM acknowledges the hospitality of the University of Rwanda-College of Science and Technology, where part of this work was conceptualised and completed.

\end{document}